# Accelerating Ground State Search of Spatial Photonic Ising Machines with Genetic-Simulated Annealing Hybrid Algorithm

Ze Zheng (郑泽)[1,4†*], Ruhui Ni (倪如慧)[2†], Jingyi Zhao (赵憬怡)[2†], Xiaojian Hu (胡小俭)[5], Wen Jiang (江文)[5], Yuegang Li (李月刚)[1], Hang Xu (徐航) [1], Tailong Xiao (肖太龙)[1,3,4,5*] and Guihua Zeng (曾贵华)[1,3,4,5]

[1] Institute for Quantum Sensing and Information Processing, State Key Laboratory of Photonics and Communications, Shanghai Jiao Tong University, Shanghai 200240, China

[2] Global College, Shanghai Jiao Tong University, Shanghai 200240, China

[3]Shanghai Research Center for Quantum Sciences, Shanghai 201315, China

[4]Hefei National Laboratory, Hefei 230088, Anhui, China

[5]Shanghai Quantum Intelligence Sensing Technology Co., Ltd, Shanghai 200240, China

[†]These authors contributed equally to this work.

[*]Corresponding author. Email: tailong_shaw@sjtu.edu.cn

Spatial photonic Ising machines (SPIMs) based on spatial light modulators (SLMs) have emerged as highly effective solvers for many tasks, including combinatorial optimization problems and spin-glass simulations. However, traditional SPIMs relying solely on the simulated annealing algorithm require a large number of measurement-feedback iterations to find a relatively optimal solution in complex energy landscapes, suffering from slow convergence and high time cost. Here, we propose an optical genetic-simulated annealing hybrid algorithm to accelerate the ground-state search of SPIMs. GA conducts a global coarse-grained search in the early iteration stage, while SA performs fine-grained local refinement in the late stage. Numerical simulations show that our method enables a higher solution quality of full-rank Max-Cut problems than pure GA or SA at different scales. We also experimentally demonstrate its superiority over conventional algorithms on a gauge-transformation time-division multiplexing SPIM for high-rank optimization problems under the same iteration budget. Our approach can be further developed with other advanced metaheuristic algorithms toward intelligent optical Ising computing systems.

---

**Introduction**

Traditional von Neumann computers can hardly obtain the optimal solutions of NP-hard problems [1, 2], such as complex materials system simulations [3-8], combinatorial optimization problems [9-13], and compressed sensing reconstruction [14-17], as the solution space of such problems grows exponentially with the number of variables [18]. Spatial photonic Ising machines (SPIMs) [19] based on spatial light modulators (SLMs) are a newly developed non-von Neumann optical computing platform to search the ground states of most NP-hard problems that have Ising formulations [20]. Generally, the target problem is modeled into an Ising model, followed by custom encoding into a light field via SLMs. The optimal solution is guided in evolution by the intensity decrease at the central point of the SLM Fourier plane. SPIMs exploit the properties of coherent lasers to directly encode large-scale fully interacting Ising models, which remains a severe scaling bottleneck for other types of Ising machines [21-24]. Recent SPIMs research has also integrated space-division [12, 25, 26], time-division [9], and wavelength-division multiplexing techniques [27], along with spin product encoding [28] and incoherent amplitude modulation schemes [29], to overcome the barriers in encoding and solving high-dimensional/rank Ising-type problems.

In the search process, SPIMs mainly rely on two categories of algorithms for ground-state search. The first category is annealing-based methods: Metropolis-based simulated annealing (SA) [27] with single-spin flipping is widely used for small-scale Ising models, while dynamic-cluster-flipping assisted SA (DSA) [30] has been developed to boost feasible solution generation for large-scale problems. However, SA-based iterations are usually stuck in local minima and unable to reach the global minimum. Adiabatic annealing schemes [31, 32] have also been demonstrated, but they all require a large number of iterations to close to the optimal solution. A common limitation of these annealing-based algorithms is that their solution quality improves slowly with increasing iterations, leading to a fundamental trade-off between convergence speed and solution quality. On the other hand, evolutionary algorithms such as particle swarm optimization (PSO) [33, 34] and genetic algorithm (GA) [35-37] have shown advantages in solving complex optimization problems on electronic computers, yet their deployment in SPIM systems remains unexplored.

In this work, we propose a hierarchical genetic-simulated annealing hybrid algorithm to accelerate the ground-state search of SPIMs, which enables rapid convergence and a high-quality solution. In the simulation, we show that the hybrid algorithm achieves solutions of higher quality than the pure GA and SA algorithms of the full-rank Max-cut problem with 50-500 spins. Combined with the gauge transformation, we introduce a time-division multiplexing SPIM using a single SLM (GT-SPIM), which avoids optical misalignment between double SLMs. To relieve the iteration rate limitations of SPIMs, we propose two solving modes: “finite iteration” and “long iteration”. We implement the GA algorithm on this all-optical computing platform to escape local minima in the early iterations, followed by the SA algorithm to search for high-quality solutions in detail. Compared to existing mainstream algorithms, we experimentally demonstrate that our proposed hybrid algorithm improves the quality of solutions of high-rank Max-cut problems with 100, 529, and 1024 spins in both modes on the GT-SPIM, respectively. The proposed approach can be further applied to other SPIM schemes or modes to enhance their performance.

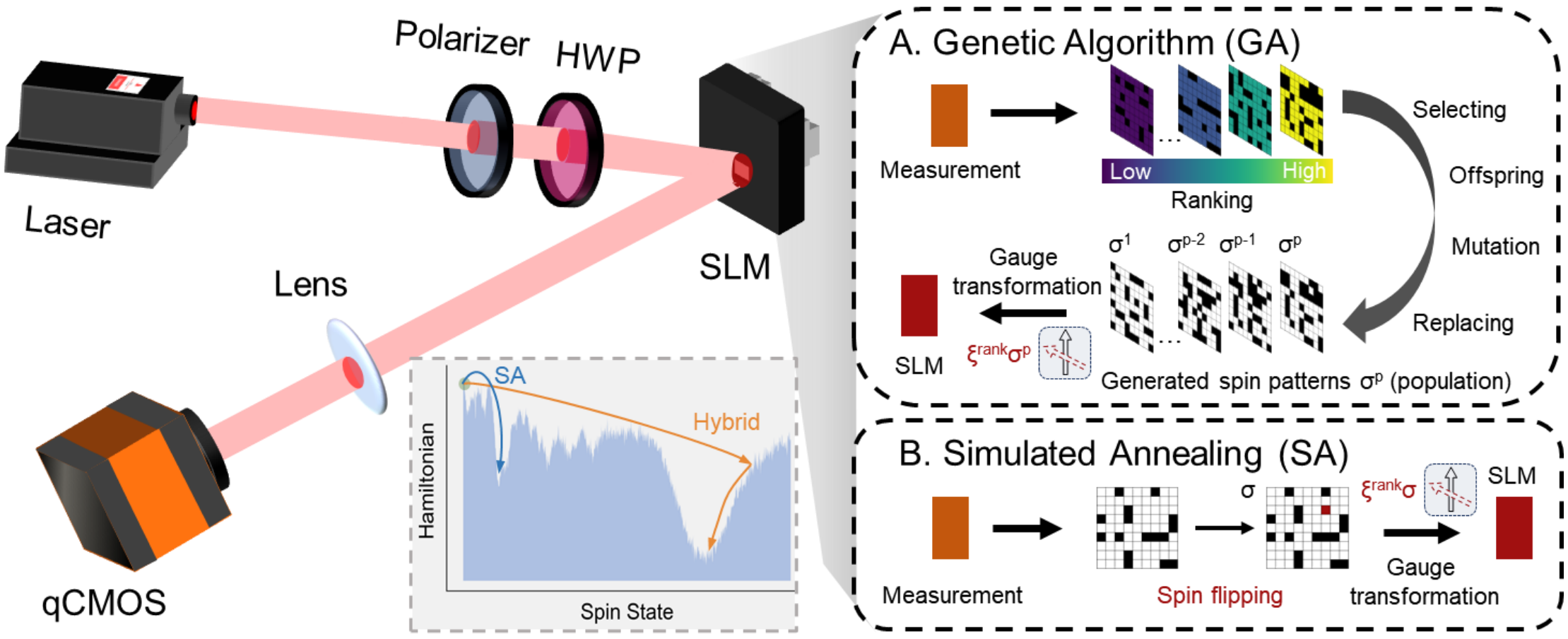

Figure 1. Concept and principle of the proposed GA-SA hybrid algorithm. The experimental setup is shown on the left. The expanded and collimated laser beam illuminates the phase-only SLM. The reflected light undergoes a Fourier transformation through a lens. The central intensity of the back Fourier plane is captured by a qCMOS camera. As shown on the right, in GA iterations, a population of $P$ spin configurations is initially generated, followed by reshaping as square patterns. After the gauge transformation, these patterns are uploaded to the SLM successively. The Ising Hamiltonian of each spin configuration is equal to the weighted summation of measurements based on different component coefficients. The new population is generated through ranking the Hamiltonians, selecting offspring, mutating, and replacing to evolve to find a spin configuration that corresponds to a lower Hamiltonian. Conducting a large-scale search with GA in the early stages helps avoid

local minima and creates an initial state that is easy for SA to anneal. In SA iterations, only one spin configuration is annealed to search for the ground truth. After measurement and weighted summation, a spin is randomly selected and flipped to generate a new spin configuration. After gauge transformation and reshaping, the patterns are uploaded to SLM for measurement. The Metropolis criterion is used to determine whether to accept this flipping.

---

**Methods**

The Hamiltonian function of an Ising model without external fields is considered as:

$$H = -\sum_{i,j} J_{i,j}\sigma_i\sigma_j, \tag{1}$$

where $J_{i,j}$ is the interaction between the i-th Ising spin $\sigma_i$ and the j-th Ising spin $\sigma_j$. The binary spin values 1 or -1. The ground state (i.e., optimal solution) of an Ising model is a spin configuration that corresponds to the lowest Hamiltonian values. By introducing eigenvalue decomposition, the Ising Hamiltonian is expressed as:

$$H = -\sum_{r=1}^{R} \lambda_r \sum_{i,j} \xi_i^r \xi_j^r \sigma_i\sigma_j, \tag{2}$$

where $r$ is the index of the r-th component, $R$ is the rank of the interaction matrix $J$, $\lambda_r$ is the coefficient (i.e., weight) of the r-th component, and $\xi^r$ is the decomposed vector related to the matrix $J$. The traditional SPIM encodes the vectors $\xi^r$ and $\sigma$ through an amplitude-type SLM and a phase-type SLM, respectively. The central intensity $I^r(0,0)$ of the far-field indicates:

$$I^r(0,0) = \sum_{i,j} \xi_i^r \xi_j^r \sigma_i\sigma_j. \tag{3}$$

The intensities of different components are measured sequentially; therefore, the corresponding Hamiltonian can be expressed as the weighted summation of the measured intensities:

$$H = -\sum_{r=1}^{R} \lambda_r I^r(0,0). \tag{4}$$

However, such schemes would introduce misalignment error between double SLMs into the SPIM system. In order to eliminate this error, we combine the gauge transformation [5] to encode $(\xi^r\sigma)_i$ through a single phase-type SLM, which is expressed as:

$$\varphi_i^r = \sigma_i\frac{\pi}{2} + (-1)^i\alpha_i^r, \tag{5}$$

where $\varphi_i$ is the modulated phase of the i-th pixel of the SLM, and $\alpha_i$ is the auxiliary parameter as:

$$\alpha_i^r = \arccos \xi_i^r. \tag{6}$$

Figure 1 presents the schematic diagram of GT-SPIM and the concept of the proposed GA-SA hybrid algorithm. During the GA iteration process, a population of $P$ spin configurations $\sigma^p$ is initially generated, followed by gauge transformation and dynamic projection-measurement by the GT-SPIM. The target functions (i.e., Hamiltonians) of each spin configuration are defined below:

$$H^p = -\sum_{r=1}^{R} \lambda_r \sum_{i,j} \xi_i^r \xi_j^r \sigma_i^p \sigma_j^p = -\sum_{r=1}^{R} \lambda_r I^{r,p}(0,0), \tag{7}$$

where $p$ is the index of the p-th spin configuration of this generation population. The population should be turned over in the next generation, similar to the survival of the fittest in nature. Each $\sigma^p$ is assigned a rank $k$ ($k$ = 1, 2, ..., $P$-1, $P$) in ascending order based on its corresponding Hamiltonian, and has a probability of $\frac{2(P-k+1)}{P(P+1)} \times 100\%$ of being selected as the parent $pa$ or $ma$. $M$ offspring would be evolved by a single-point crossover strategy [35], which indicates:

$$Offspring = pa \cdot T + ma(1-T), \tag{8}$$

where $T$ is the random coordinate of a specific spin. Each spin of the offspring also has a probability of $q$ to be mutated (from 1 to -1 or from -1 to 1), which is similar to the gene mutation process in natural evolution. The eventually generated $M$ new offspring will replace the $M$ lowest-ranked spin configurations from the previous generation, resulting in a new population of $P$ spin configurations, followed by gauge transformation and uploading to the SLM again.

The GA search process, which accounts for $g$ percent of the total number of iterations, performs an early-stage broad search to escape local optima and prepare a favorable initial state for the SA search phase. SA is suitable for small-scale and precise ground state search by flipping one spin randomly in each iteration. During the SA search process, the next-generation spin configuration $s'$ is generated by randomly flipping a spin of the last-generation spin configuration $s$ per iteration. After gauge transformation, the corresponding intensities of different components are measured, followed by calculating the Hamiltonian, as in Eqs. (3) and (4). The Metropolis criterion, as shown in Eq. (9), is used to determine the probability of accepting this new spin configuration.

$$Prob. = \begin{cases} 1, & if\ \Delta H < 0 \\ e^{\frac{-\Delta H}{T}}, & if\ \Delta H \geq 0 \end{cases}, \tag{9}$$

where $\Delta H = H(s') - H(s)$ is the variation of the Hamiltonian, and $T$ is a temperature parameter, which decreases with the increase of iteration count in a predefined manner (i.e., annealing). The hierarchical fusion of GA and SA would accelerate ground state search by escaping local minima for various-scale complex problems.

**Simulations**

We first demonstrate the effectiveness of our approach by comparing it with the pure GA and SA algorithms in solving the full rank Max-cut problems. The max-cut problem is a classic NP-hard optimization problem in graph theory [38], which aims to partition the set of vertices of a graph into two disjoint subsets such that the total weight of the edges connecting the two subsets (i.e., the "cutting edges") is maximized. The target function *Val* of the Max-cut problem is shown below [21]:

$$Val = \sum_{i,j} w_{ij} \frac{1-\sigma_i\sigma_j}{2} = \frac{1}{4}\sum_{i,j} w_{ij} - \frac{1}{4}\sum_{i,j} w_{ij}\sigma_i\sigma_j, \tag{10}$$

where $w_{i,j}$ is the connection weight between the i-th vertex $\sigma_i$ and the j-th vertex $\sigma_j$, and $\sigma \in \{-1, 1\}$. Vertices with the same value belong to the same set after cutting. $\sum_{i,j} w_{ij}$ is a constant to a specific graph, and maximizing *Val* is equivalent to minimizing $\sum_{i,j} w_{ij}\sigma_i\sigma_j$. By setting $J_{i,j} = -w_{i,j}$, we can map this problem into the Ising formulation as Eq. (1), followed by solving it through GT-SPIM.

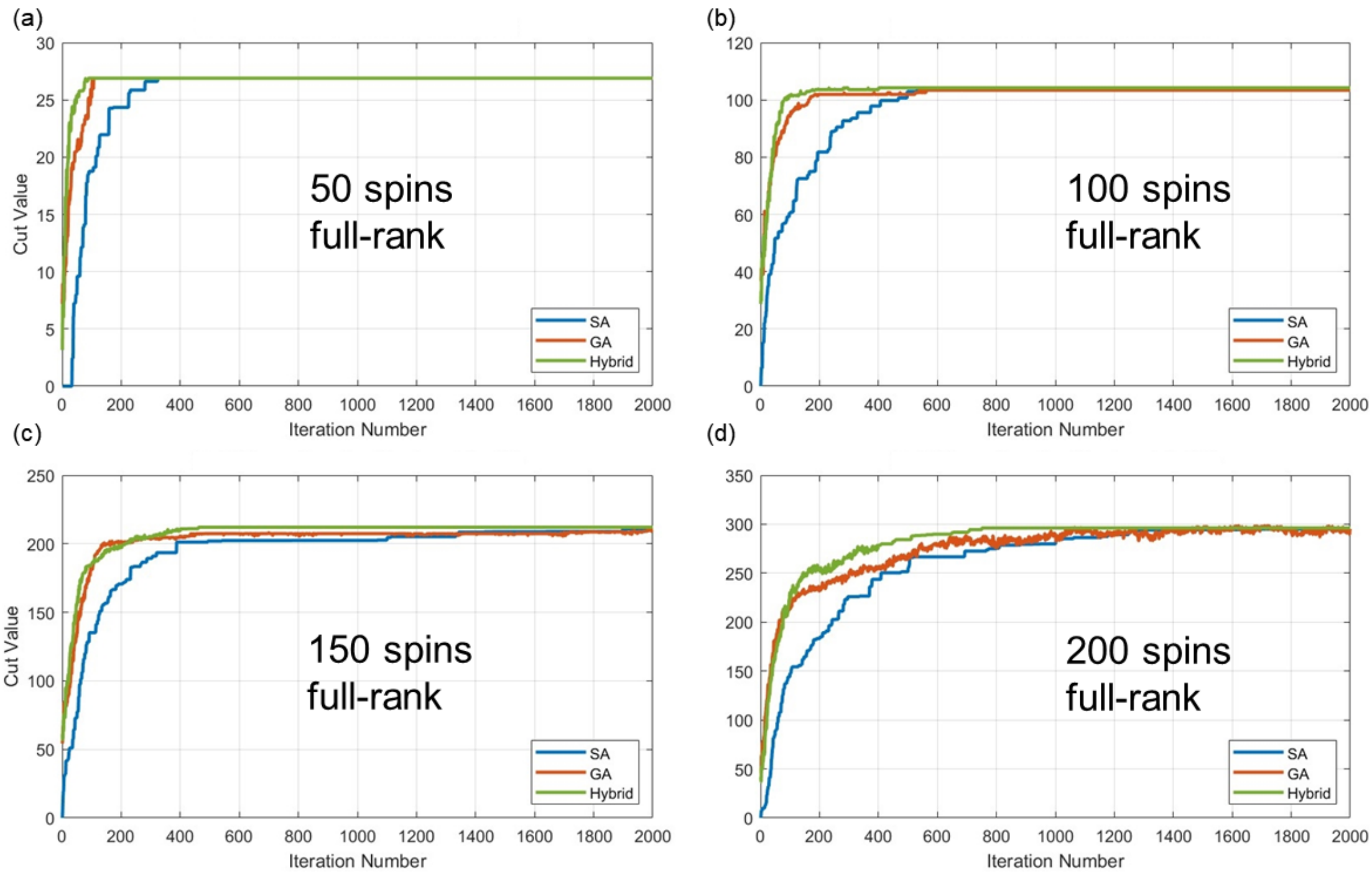

Figure 2. Comparison of the curves of cut values with the number of iterations with the pure SA (in blue), the pure GA (in orange), and our proposed hybrid algorithm (in green), for the full-rank Max-cut problems of small sizes: (a) 50 spins, (b) 100 spins, (c) 150 spins, and (d) 200 spins.

---

In simulation, we randomly generate the full-rank $J$ matrices for small spin-size (i.e., 50, 100, 150, and 200 spins), and large spin-size (i.e., 300 and 500). We set the proportion of GA iterations to the total number of iterations as $g = 10\%$, the population size as $P = 50$, the number of new spin configurations generated by crossover per generation as $M = 35$, and each spin has a $q = 0.5\%$ probability of mutation. In SA iteration, we set the initial temperature $T_0$ = 10 and adopt a logarithmic strategy for cooling, as $T = \frac{T_0}{ln\,(iter+1)}$, where *iter* is the iteration number. The total iterations are set to 2000 for small spin-scales, and 5000 for large spin-scales, respectively. The initial spin configuration is set to an all-one vector. The iterative processes of the pure SA, the pure GA, and the proposed GA-SA hybrid algorithm for solving small spin sizes and large spin sizes are shown in Figure 2 and Figure 3, respectively. Tables 1 and 2 present the maximum cut values found by these three algorithms under different spin sizes and the corresponding number of iterations.

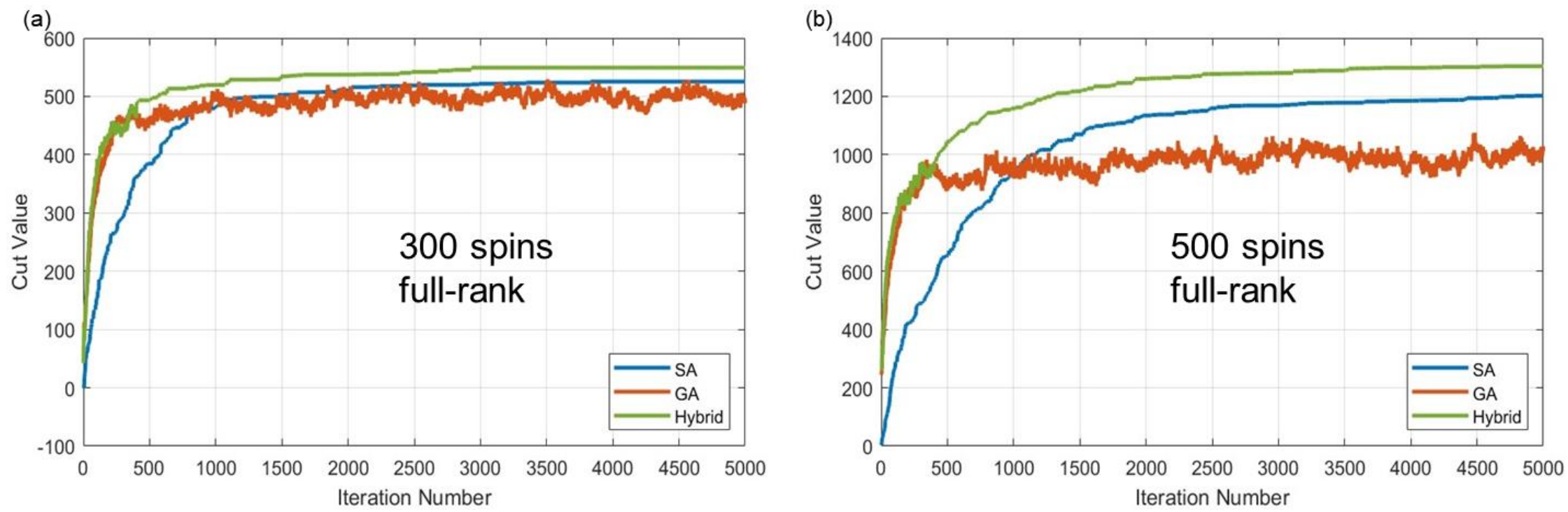

Figure 3. Comparison of the curves of cut values with the number of iterations with the pure SA (in blue), the pure GA (in orange), and our proposed hybrid algorithm (in green), for the full-rank Max-cut problems of small sizes: (a) 300 spins, and (b) 500 spins.

---

Table 1. The Max-cut values and the iteration steps to reach the maximum of the SA, GA, and our approach for 50 spins, 100 spins, and 150 spins.

| | N=50 | | N=100 | | N=150 | |
|---|---|---|---|---|---|---|
| Algorithm | Max-cut Value↑ | Steps to the maximum↓ | Max-cut Value↑ | Steps to the maximum↓ | Max-cut Value↑ | Steps to the maximum↓ |
| SA | **26.89** | 324 | 103.84 | 544 | 210.86 | 1908 |
| GA | **26.89** | 102 | 103.82 | 682 | 211.03 | 1986 |
| Hybrid (ours) | **26.89** | **79** | **104.22** | **232** | **212.12** | **462** |

Table 2. The Max-cut values and the iteration steps to reach the maximum of the SA, GA, and our approach for 200 spins, 300 spins, and 500 spins.

| | N=200 | | N=300 | | N=500 | |
|---|---|---|---|---|---|---|
| Algorithm | Max-cut Value↑ | Steps to the maximum↓ | Max-cut Value↑ | Steps to the maximum↓ | Max-cut Value↑ | Steps to the maximum↓ |
| SA | 294.88 | 1662 | 525.23 | 3830 | 1202.65 | 4998 |
| GA | **298.26** | 1613 | 529.38 | 3507 | 1075.61 | **4484** |
| Hybrid (ours) | 296.23 | **751** | **549.19** | **2932** | **1303.39** | 4692 |

As illustrated in Fig. 2 and Fig. 3, our proposed hybrid algorithm demonstrates a more pronounced increase in the Max-cut values in the early stages, supported by GA. In the subsequent stages, with the support of the SA algorithm, the hybrid algorithm converges smoothly to a high-quality solution, in contrast to the severe oscillations exhibited by GA. These advantages become particularly evident in large-scale simulations and long iteration conditions. As shown in Tables 1 and Table 2, compared to pure SA and GA, the hybrid algorithm consistently finds higher-quality solutions in most tests with fewer iteration steps.

**Experiments**

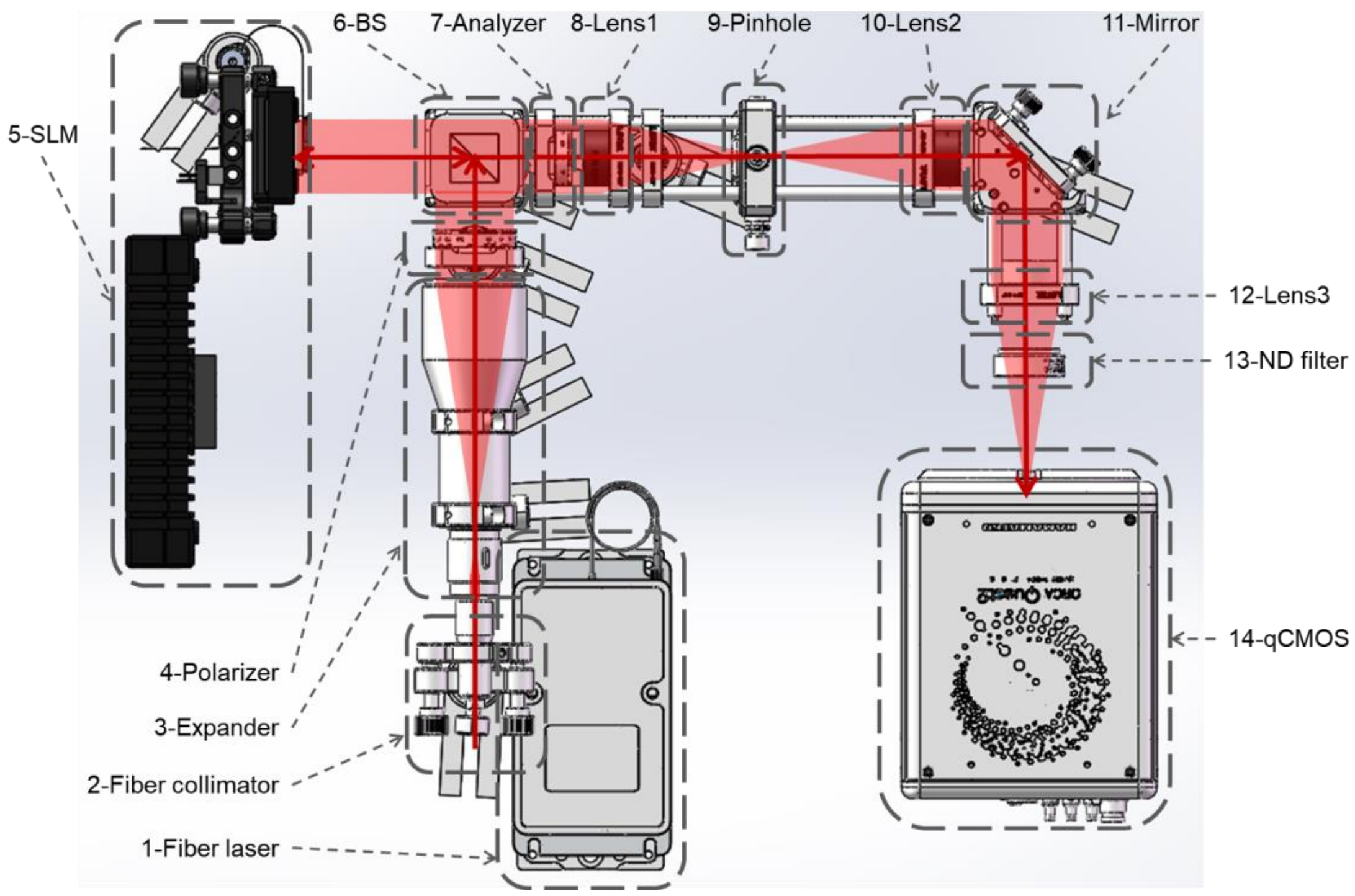

Figure 4. Experimental setup. A 780 nm continuous laser beam (in red) emitted from a fiber laser (1, PL-DFB-780-D-1-PA, Microphotons) is first collimated by a fiber collimator (2, GCX-L30APC-780, DHC), then expanded by a beam expander (3, FBE-10X-B, LBTEK) and purified in linear polarization state by a polarizer (4, OPPF1-NIR, JCOPTIX). The polarized beam is incident onto a beam splitter (6, GCC-403112, DHC), which transmits the beam to a phase-type SLM (5, GAEA-2.1-NIR-069B, HOLOEYE) for phase modulation. The phase-modulated beam reflected from the SLM is redirected by the same beam splitter (6) to a polarization analyzer (7, OPPF1-NIR, JCOPTIX), which is aligned with the polarization direction parallel to the polarizer (4) and filters the light. The modulated beam passes through a 4f system consisting of Lens1 (8, f = 75 mm, AC254-075-B-ML, Thorlabs), a 300 μm pinhole (9, PS1-300H, LBTEK) placed at the Fourier plane for noise filtering, and Lens2 (10, f = 75 mm, AC254-075-B-ML, Thorlabs). The filtered beam is then folded by a mirror (11, BDM1E-B, LBTEK), focused by Lens3 (12, f = 100 mm, AC254-100-B-ML, Thorlabs), and adjusted in intensity by a neutral density filter (13, OD = 3.0, OFR1-30M, JCOPTIX) before being detected by a quantitative complementary metal-oxide-semiconductor (qCMOS) camera (14, C15550-22UP, Hamamatsu) for intensity measurement.

---

The experimental setup is shown in Fig. 4, and we perform the GT-SPIM scheme on this system. We encode each spin into a super-pixel with 4-by-4 pixels of the employed SLM. The iteration rate of GT-SPIM, which is 50 Hz in our experiments, is limited by the refresh rate of the SLM. In order to relieve this limitation and move towards practical applications, we propose and test different algorithms under two solving modes, named “finite iteration” and “long iteration”. The "finite iteration" mode necessitates algorithms that can rapidly converge toward the optimal solution. In this study, we compared our approach with other algorithms to assess the solution quality for solving the 100-spin Max-cut problem after 30 and 800 iterations, as depicted in Fig. 5. The "long iteration" mode necessitates algorithms that can enhance solution quality through multiple iterations. In this study, we compared our approach with other algorithms in terms of solution quality after 10,000 iterations for solving the 529-spin and 1024-spin Max-cut problem, as illustrated in Fig. 6.

In the experiment, for the GA process, we set the population size $P = 10$ and replace half of the spin configurations in the population at each generation to reduce the total runtime of the system. Each spin has a 0.5% probability of undergoing mutation. For the SA process, we set the initial temperature $T_0 = 10$ and employ the linear annealing strategy $T = \beta T_0$ to expedite the annealing rate, where β is a constant coefficient, which is 0.7 in "finite iteration" mode and 0.999 in "long iteration" mode. For a more general comparison, we also experimentally considered DSA [30], which randomly flips a small dynamic cluster of spins at each iteration. In our experiments, the upper limit on the number of flipped spins was set to 10% of the spin size. We further proposed and compared the DAS-M algorithm, which combines 20% DSA search with 80% SA search. This ratio is the same as our hybrid algorithm (i.e., 20% GA and 80% SA).

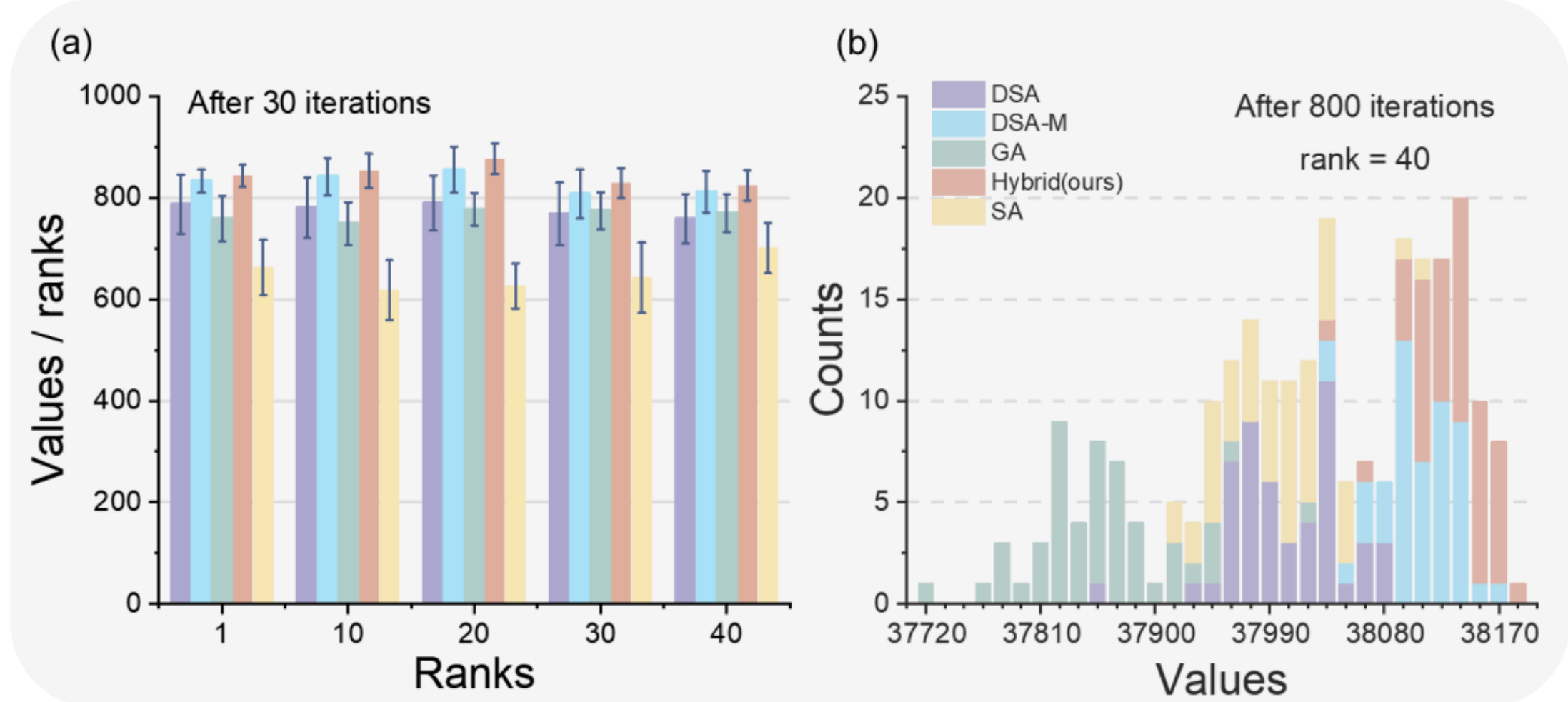

Figure 5. Performance of different algorithms under the "finite iterations" mode in solving the 100-spin Max-cut problem. Each experiment is run in parallel 50 times to calculate the mean and standard deviation (SD) for statistical analysis. (a) The bar chart shows the mean Max-cut values searched by different algorithms after 30 iterations when the ranks are 1, 10, 20, 30, and 40, respectively. The vertical coordinates are converted to the same order of magnitude by dividing the corresponding ranks. The error bars represent SDs. (b) This stacked chart shows the distribution of solutions searched by different algorithms after 800 iterations for the 100-spin Max-cut problem with a rank of 40.

---

We customize interaction matrices $J^R$ of different ranks $R$ through the inverse process of eigenvalue decomposition, as shown below:

$$J^R = -\sum_{r=1}^{R} \sum_{i,j} \xi_i^r \xi_j^r, \tag{11}$$

where $\xi$ is a $1 \times N$ vector. Each element of $\xi$ is randomly selected from the set {-1, 0, 1, 2} in the "finite iterations" mode, and from the set {0, 1} in the "long iterations" mode. $N$ is the number of spins. The $J^R$ matrix has the desired rank. Although the elements on its main diagonal (i.e., self-interactions) are non-zero, this only adds a constant to the measurements and does not alter the manner in which vertices are partitioned [27]. The displayed Max-cut values are calculated by setting the elements on the main diagonal of $J^R$ matrices to 0.

Figure 5 demonstrates the performance of different algorithms under the "finite iterations" mode. All spins are encoded into the square region of $N = 10 \times 10$ super-pixels at the center of the SLM. The initial spin configuration is fairly set to an all-one vector. After 30 iterations, our hybrid algorithm (in red) performs best on the mean value with the smallest SDs across the ranks, followed by the DSA-M (in blue) algorithm, as shown in Fig. 5(a). After 800 iterations, the results from our hybrid algorithm's 50 runs concentrated around a relatively large Max-cut value, followed by those from DSA-M, as shown in Fig. 5(b). As the number of iterations increased, the performance of the SA (in yellow) exhibited a marked improvement, approaching that of the DSA (in purple) by the 800th iteration. The GA (in green) is distinguished by its rapid search speed on a large spin-size; however, it faces challenges in converging on an optimal solution. This also indicates that our proposed method has the potential to expedite the search for the ground state of the Ising problem over other algorithms within a limited timeframe.

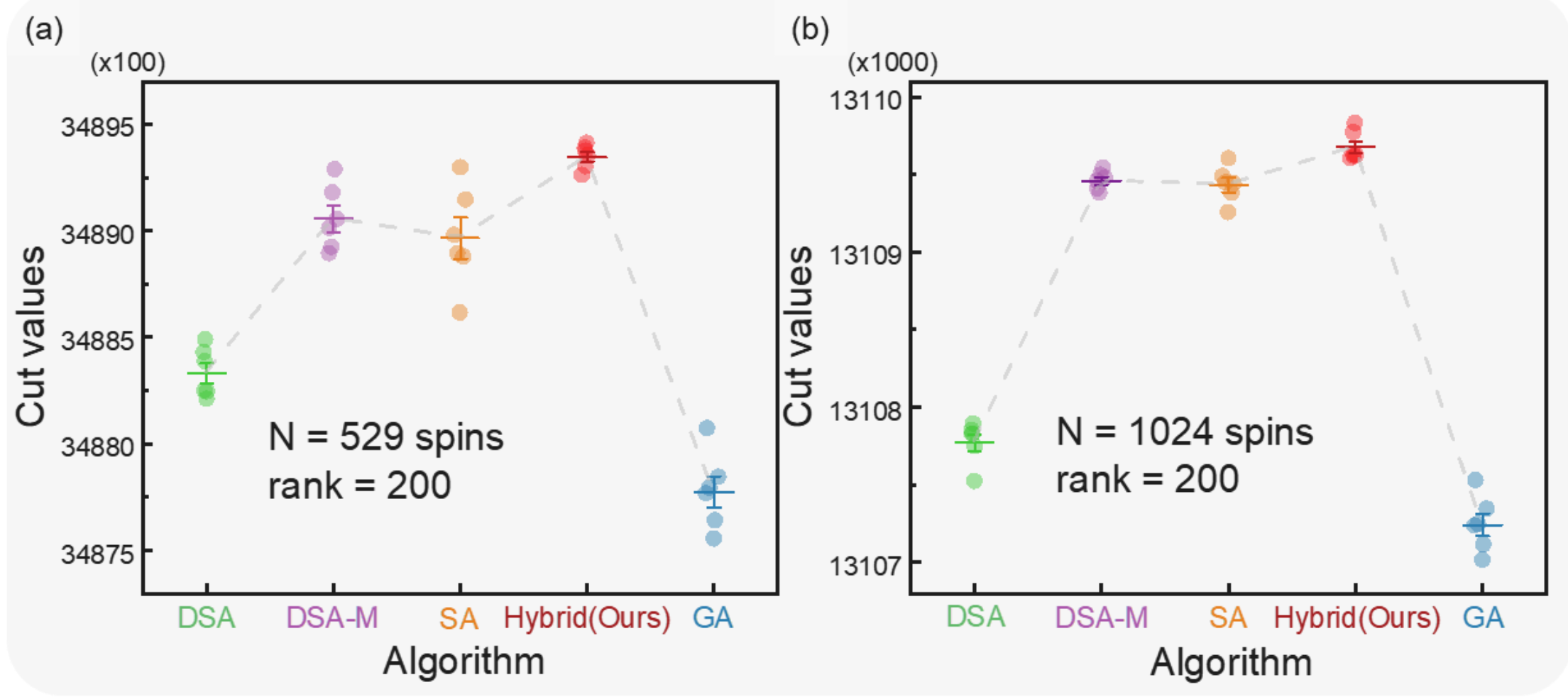

Figure 6. Performance of different algorithms under the "long iterations" mode in solving (a) 529-spin, and (b) 1024-spin Max-cut problems. Each algorithm is executed 6 times for each problem with 10,000 iterations, and the corresponding results are shown as circular data points. Error bars of different colors represent the mean values and SDs for the corresponding algorithm. A gray dashed line connects these mean values.

---

Table 3. The best and mean Max-cut values of different algorithms for 529-spin and 1024 spins.

| | Best↑ (529 spins) | Mean↑ (529 spins) | Best↑ (1024 spins) | Mean↑ (1024 spins) |
|---|---|---|---|---|
| Best-known | 3,501,441 | / | 13,130,222 | / |
| Hybrid (ours) | **3,489,410** | **3,489,347** | **13,109,829** | **13,109,678** |
| SA | <u>3,489,295</u> | 3,488,966 | <u>13,109,601</u> | 13,109,451 |
| DSA-M | 3,489,286 | <u>3,489,056</u> | 13,109,541 | <u>13,109,459</u> |
| DSA | 3,888,486 | 3,488,332 | 13,107,888 | 13,107,773 |
| GA | 3,488,069 | 3,487,775 | 13,107,523 | 13,107,241 |
| GW-SDP | 3,488,973 | / | 13,109,187 | / |

Figure 6 demonstrates the performance of different algorithm under the "long iterations" mode with 10,000 iterations for solving $N = 23 \times 23 = 529$ spins, and $N = 32 \times 32 = 1024$ spins Max-cut problems. The initial spin configuration is set to a random vector composed of 1 and -1. The interaction matrix $J^R$ is custom-defined by Eq. (11), where the elements of $\xi$ are randomly selected from the set {0, 1}. We make the ranks $R = 200$ for $J^R$ of both 529-spin and 1024-spin problems. As shown in Table 3, the best and mean values of the various algorithms from each of the six parallel runs are summarized. The best-known values are solved by the academic version of Gurobi, and the GW-SDP algorithm can achieve an expected value that is greater than 87.8% of the optimal solution for Max-cut problems. As illustrated in Fig. 6 and Table 3, the hybrid algorithm generated Max-cut values that are close to the best-known values for both tasks, delivering higher-quality average performance and enhanced stability. This characteristic not only facilitates the solution of tasks when sufficient computational resources are available but also provides a foundation for further development of other SPIM systems to pursue higher-quality solutions.

## Discussion

In summary, we have proposed and demonstrated the hierarchical genetic-simulated annealing hybrid algorithm to accelerate ground state search of SPIMs. We have also improved the time-division multiplexing SPIM by incorporating gauge transformation (i.e., GT-SPIM) and applied optical GA to this all-optical computing platform. This enables the encoding and solving of high-rank Ising problems using only a single phase-type SLM. By comparing with other algorithms, we have demonstrated the effectiveness of the proposed hybrid algorithm through simulations and experiments conducted on the GT-SPIM. We propose two operating modes (the "finite iterations" and the "long iterations") for GT-SPIM, which are respectively applied to two application scenarios: time-limited accelerated search and high-precision ground state search. The proposed algorithm demonstrates high potential for efficient practical application and is expected to assist other SPIMs in enhancing their success probabilities in solution implementation and boosting broad applications, including Boltzmann machines [39] and extreme learning machine [40].

It should be noted that the major limitation of our approach is the SLMs' update rate. The time-division multiplexing scheme achieves an interaction matrix with a coding rank greater than 1 at the cost of time resources. Under the current experimental conditions, it consumes a considerable amount of time. The recently developed DMD encoding scheme [29] and parallel search scheme [12, 41] are helpful in alleviating this limitation. The significant progress in SLMs with a GHz refresh rate [42, 43] is expected to be applied to SPIMs in the near future, thereby significantly improving this predicament. The incorporation of strategies such as adaptive parameters and multi-objective optimization has the potential to further enhance the proposed hybrid algorithm's ability to search the ground state.

## Acknowledgments

This work was supported by the National Key R&D Program of China (No. 2025YFF0515504), the National Natural Science Foundation of China (No. 62401359), the State Key Laboratory of Photonics and Communications, the Quantum Science and Technology - National Science and Technology Major Project (No. 2021ZD0300703), the Shanghai Municipal Science and Technology Major Project (No. 2019SHZDZX01), and the Participation in Research Program (PRP) of Shanghai Jiao Tong University (No. T030PRP47075).

## Author Contributions

G.Z. conceived the research project. Z.Z. designed the scheme. R.N. implemented the algorithms. R.N. and J.Z developed the simulation and analyzed the simulation data. Z.Z. and X.H. performed the experiments. G.Z. and T.X. supervised the project. Z.Z. analyzed the results and prepared the manuscript. All the authors participated in the discussion and confirmed the final manuscript.

## Conflicts of Interest

The authors declare no conflicts of interest.

## Data Availability Statement

The data that support the findings of this study are available from the corresponding author upon reasonable request